\documentclass[12pt]{iopart}

\usepackage{iopams}
\usepackage{graphicx}
\usepackage{bm}
\newcommand{\be}{\begin{equation}}\newcommand{\ee}{\end{equation}}
\newcommand{\bea}{\begin{eqnarray} }\newcommand{\eea}{\end{eqnarray}}
\newcommand{\beaa}{\begin{eqnarray} }\newcommand{\eeaa}{\end{eqnarray}}
\newcommand{\bsa}{\begin{subeqnarray}}\newcommand{\esa}{\end{subeqnarray}}
\newcommand{\ba}{\begin{array}}\newcommand{\ea}{\end{array}}
\newcommand{\bit}{\begin{itemize}}\newcommand{\eit}{\end{itemize}}
\newcommand{\ben}{\begin{enumerate}}\newcommand{\een}{\end{enumerate}}

\def\bol#1{\mbox{\boldmath\footnotesize $#1$\normalsize\unboldmath}}

\def\1{{_{1}}}\def\2{{_{2}}}
\def\Real{\Re}
\def\Imag{\Im}
\def\vect#1{{\bm #1}}

\def\bol#1{\mbox{\boldmath\footnotesize $#1$\normalsize\unboldmath}}

\begin{document}

\title[]{Nambu--Goldstone dynamics and generalized coherent-state functional integrals}

\author{Massimo Blasone$^1$\footnote{Corresponding
author: blasone@sa.infn.it} and Petr
Jizba$^2$\footnote{Corresponding author: p.jizba@fjfi.cvut.cz}}

\address{$^1$ INFN, Gruppo Collegato di Salerno and Universita' di Salerno,
Via Ponte don Melillo, 84084 Fisciano (SA), Italy}
\address{$^2$ FNSPE, Czech Technical
University in Prague, B\v{r}ehov\'{a} 7, 115 19 Praha 1, Czech Republic}


\begin{abstract}
The present paper gives a new method of attack on the Nambu--Goldstone dynamics
in spontaneously broken theories. Since the target space of the Nambu--Goldstone fields
is a group coset space, their effective quantum dynamics can be naturally phrased
in terms of generalized coherent-state functional integrals.
As an explicit example of this line of reasoning we construct a low-energy effective Lagrangian for
the Heisenberg ferromagnet in broken phase. The leading field configuration in
the WKB approximation leads to the Landau--Lifshitz equation for quantum
ferromagnet. The corresponding linearized equations allow to identify the Nambu--Goldstone boson
with ferromagnetic magnon.
\end{abstract}

\vspace{2pc}
\noindent{\it Keywords}: Nambu--Goldstone bosons, Coherent-state functional integrals,
Non-linear $\sigma$ models, Landau--Lifshitz magnons

\maketitle

\section{Introduction \label{intro}}

Functional integrals provide indisputably
a powerful tool in diverse areas of physics,
both computationally and conceptually. They often offer
the easiest route to derivation of perturbation expansions, accommodate
naturally gauge symmetry  and serve as an excellent framework for non-perturbative
analysis~\cite{Kleinert,Zinn}. Growing popularity among practitioners in both high-energy and
solid-state physics enjoy functional integrals which are based on
the occupation number representation or on the Fock space.
In contrast, the functional integrals that are rooted in the over-complete set of coherent states (CS)
are used comparatively less. Despite their cleaner mathematical structure
are the CS-based functional integrals still rather interesting curiosity
than full-fledged tools of particle or solid-states physics.

It is purpose of this paper to call attention to the fact that CS-based functional
integrals constructed from the so-called  group-related
or generalized CS~\cite{Perelomov:1986tf,Arecchi:1972a,DAriano:1985b,DAriano:1985a,Rasetti:1975a}
%
offer
a very natural tool in theory of critical phenomena with genuine phenomenological
implications. In particular, they have a built-in quality to describe
an effective low-energy behavior of systems with spontaneous
breakdown of a global continuous symmetry provided the interest lies in the
low-energy gapless excitations known as Nambu--Goldstone (NG) bosons.
We will illustrate our point by employing the generalized CS functional integrals
to investigate the low-energy behavior of ferromagnets in the broken phase,
i.e., below the Curie
temperature.

The structure of the paper is as follows.
To set the stage we recall in the next section some
fundamentals of the group-related CS with a special emphasis on the $SU(2)$ CS.
Section~\ref{ApJ} is devoted to formulation of functional integrals by means of generalized CS.
A natural appearance of the geometric Berry--Anandan phase in the action of the CS functional integrals and
the way how it may affect the dynamics is also discussed. As an explicit example we derive the $SU(2)$ CS functional integral. The r\^{o}le of the group quotient space as an arena for the dynamics of NG fields is discussed in Section~\ref{ApJ0}. There we also prove the NG theorem with the help of the coset-space construction of spontaneous symmetry breakdown (SSB).
Distinction between relativistic and non-relativistic versions of the NG theorem is stressed.
In Section~\ref{ApK3}  we observe
that transition amplitudes as well as partition function for NG modes can be formulated via
the generalized CS functional integrals.  To put more flesh on the bare bones we investigate
the low-temperature properties of the quantum Heisenberg model of ferromagnet
in a broken phase. The corresponding CS functional integral can be identified with
the $SU(2)/U(1)\!-\!\sigma$ model. The WKB approximation yields in the limit of continuous spin lattice (i.e., large wavelength limit) Landau--Lifshitz equations for quantum forromagnet. Linearized version of the latter equations
allows to identify the NG field with the massless spin wave. The NG boson then corresponds to ferromagnetic magnon.
Various remarks and generalizations are postponed to the concluding section.

\section{Group-related  coherent states \label{ApD}}

%

To construct the CS related to a Lie group $G$ we follow here Ref.~\cite{Perelomov:1986tf}.
Let $\hat{D}(g), \ g\in G$ be an
irreducible {\em unitary}
representation of $G$ acting in some Hilbert space ${\mathcal{H}}$.
We choose a normalized fiducial state vector
in ${\mathcal{H}}$ and denote it  as $|0\rangle$.
The generalized CS's
corresponding to $G$ are then defined as
\begin{eqnarray}
|0(g) \rangle \ = \ \hat{D}(g)|0\rangle \;\;\;\;\; \mbox{for}\;\;\forall
g\in G\, .\label{1.1.2.3.370}
\end{eqnarray}
With foresight of applications in the SSB theory, we have denoted the group-related CS as
$|0(g) \rangle$. Two CS $|0(g_1)\rangle$ and
$|0(g_2)\rangle$ represent the same physical state in ${\mathcal{H}}$ if
\begin{eqnarray}
\hat{D}(g_1)|0\rangle \ = \
e^{i\alpha(g_1,g_2)}\hat{D}(g_2)|0\rangle\;\;\Leftrightarrow\;\;
\hat{D}(g_2^{-1}g_1)|0\rangle \ = \ e^{i\alpha(g_1,g_2)}|0\rangle\, .
\label{1.1.2.3.37}
\end{eqnarray}
%
Defining the {\em
stability} group $H_{|0\rangle}$ as a group of transformations
leaving $|0\rangle$ invariant (up to a phase), i.e.,
\begin{eqnarray}
H_{|0\rangle} \ = \ \{h \in  G: \hat{D}(h)|0\rangle =
e^{i\beta(h)}|0\rangle\, , \beta(h)\in {\mathbb{R}}\}\, ,
\end{eqnarray}
the distinct $G$-related CS can be parameterized by elements
of the coset $G/H_{|0\rangle}$.
%
Since $H_{|0\rangle}$'s for different fiducial states are mutually
isomorphic subgroups of $G$
we will simply use $H$ instead of $H_{|0\rangle}$.

Let $d\mu(g)$ be the left-invariant group measure,
i.e., for any fixed $g_0\in G$, $d\mu(g_0\cdot g) = d\mu(g)$.
%
 Having  $d\mu(g)$, the measure on the coset space
$G/H$ is naturally induced. We denote it as $d\bol{\zeta}$. The
resolution of the unity can be then written as

\begin{eqnarray}
\hat{\mathbf{1}} \ = \ c \int_{G}d\mu(g) \ |0(g)\rangle\langle 0(g)| \ = \ c
\int_{G/H}d\bol{\zeta} \ |0(\bol{\zeta})\rangle\langle 0(\bol{\zeta})|\, .
\label{1.2.3.38}
\end{eqnarray}
Here $c$ is determined so as to fulfill the consistency condition
\begin{eqnarray}
1 \ = \ \langle 0({\bol{\zeta}}')|0({\bol{\zeta}}')\rangle \ = \ c\int_{G/H}
d{\bol{\zeta}} \ |\langle 0({\bol{\zeta}}')|0({\bol{\zeta}})\rangle|^2,\;\;\;\;\;{\bol{\zeta}}' \in G/H\, . \label{1.1.2.3.44}
\end{eqnarray}
%
It is thus meaningful to restrict oneself to
representations $\hat{D}(g)$  that are square integrable over the quotient $G/H$.
More up-to-date view on the group-related CS
together with much of the background material can be found, for instance, in
Refs.~\cite{GazeauI,GazeauII}.

\subsection{SU(2) coherent states}

For our purpose we will specifically consider the $SU(2)$ CS.
The $SU(2)$\index{Group!SU(2)} group has three generators
$\hat{J}_1,\hat{J}_2, \hat{J}_3$ which close the $su(2)$
algebra
\begin{eqnarray}
[\hat{{J}}_+, \hat{{J}}_-] \ = \ 2 \hat{{J}}_3\;\;\;\;\;\;\;\;\; [\hat{{J}}_3, \hat{{J}}_{\pm}]\ =
\ \pm \hat{{J}}_{\pm}\, .
\end{eqnarray}
Here
$\hat{{J}}_{\pm} =
\hat{{J}}_1 \pm i\hat{{J}}_2$. The unitary
irreducible representations of the
$su(2)$ algebra are {\em finite-dimensional} and are spanned
by states $|j,m\rangle$ fulfilling
\begin{eqnarray}
&&\mbox{\hspace{-8mm}}\hat{{J}}_3|j,m\rangle \ = \ m  |j,m\rangle\, ,\nonumber \\[2mm]
&&\mbox{\hspace{-8mm}}\hat{{J}}_{\pm} |j,m\rangle \ = \ \sqrt{(j\mp m)(j
\pm m + 1)}\ |j,m\pm 1\rangle \, , \;\;\;\; \;(|m| \leq j)\, .
\label{1.2.2.38}
\end{eqnarray}

The representations of $SU(2)$ are labeled by the
eigenvalues of the $su(2)$ Casimir operator:
\begin{eqnarray}
\hat{{\mathcal{C}}} \ = \ \hat{{\mathbf{J}}}^2
\ = \  \mbox{$\frac{1}{2}$}(\hat{{J}}_+\hat{{J}}_- +
\hat{{J}}_-\hat{{J}}_+) + \hat{{J}}_3^2 \ = \ j(j+1)\hat{\mathbf{1}} \, ,
\end{eqnarray}
i.e.,
\begin{eqnarray}
\hat{{\mathbf{J}}}^2 |j,m\rangle \ = \ j(j+1)|j,m\rangle \;\;\;\;\;\,
\mbox{with} \;\;\; \; \; \; j = 0,\mbox{$\frac{1}{2}$}, 1,
\mbox{$\frac{3}{2}$}, \ldots \, .
\end{eqnarray}
As the fiducial vector we might choose the state $|j,-j\rangle$.
In this way each representation has its
unique fiducial state --- ``vacuum state" $|0\rangle \equiv
|j,-j\rangle$. The stability group
is the subgroup of
rotations around the $z$-axis, thus $H = U(1)$.
According to Eq.~(\ref{1.2.3.38}) the distinct CS are
labeled by ${\bol{\zeta}} \!\in  G/H$. By noting that
$SU(2)/U(1) \cong {\mathcal{S }}^2$ we can identify
${\bol{\zeta}}$ with the spherical angles $\theta$ and
$\varphi$. The associated CS can then be written as
$|0(\theta,\varphi)\rangle$:
\begin{eqnarray}
|0(\theta,\varphi)\rangle  \ = \ \hat{D}(\theta,\varphi)|0\rangle \ = \
\exp\left[i\theta ({\hat{{\bf{J}}}}\cdot {\bf{n}}) \right]\!|0\rangle \, ,
\end{eqnarray}
with the unit vector ${\mathbf n} = (\sin \varphi, \cos
\varphi, 0)$. Using the Gauss decomposition formula
\begin{eqnarray}
\hat{D}(\theta,\varphi) \ = \ e^{\xi \hat{{J}}_+}\ e^{\log(1+|\xi|^2)\hat{{J}}_3}\
e^{-\xi^* \hat{{J}}_-}, \;\;\;\;\;\;\; \xi \ = \ \tan\frac{\theta}{2}\
e^{i\varphi}\, ,
\end{eqnarray}
one can alternatively use the more economical form
\begin{eqnarray}
|0(\theta,\varphi)\rangle  \ = \ (1 + |\xi|^2)^{-j} e^{\xi
\hat{{J}}_+}|0\rangle \ \equiv \ |0(\xi)\rangle \, . \label{1.2.2.40}
\end{eqnarray}
The scalar product of two CS $|0(\xi)\rangle$ can be written in
the form
\begin{eqnarray}
\langle0({\xi'^*})|0(\xi)\rangle
\ = \ \frac{(1+ \xi'^*\xi)^{2j}}{(1 + |\xi'|^2)^{j}(1 +
|\xi|^2)^{j}}\, . \label{1.2.2.44}
\end{eqnarray}
%
%
An important implication of Eq.~(\ref{1.2.2.44}), that will be relevant later, is that
\begin{eqnarray}
|\langle0(\xi'^*)|0(\xi)\rangle|^2
\ = \ \left(\frac{1+ {\bf m}'\cdot{\bf m}}{2}\right)^{\!\!2j}\, .
\end{eqnarray}
Here ${\bf m} = (\sin\theta \cos\varphi, \sin\theta \sin\varphi,
\cos\theta)$ is the unit vector parameterizing $ {\mathcal{S}}^2$. Analogous arguments hold also for ${\bf m}'$.
Since the $SU(2)$ CS can be equally well parametrized by ${\bf m}$
we will use the notation $|0({\bf m})\rangle \equiv  |0(\xi)\rangle = |0(\theta,\varphi)\rangle$.
According to Eq. (\ref{1.2.3.38}) the resolution of the
unity reads
\begin{eqnarray}
\hat{\mathbf{1}} \ = \ \int_{SU(2)}d\mu(g) \ |0(g)\rangle\langle 0(g)| \ = \ c
\int_{{\mathcal{S}}^2}d{\bf m} \ |0({\bf m})\rangle\langle 0({\bf
m})|\, .
\end{eqnarray}
The constant $c$ is determined from the normalization condition
\begin{eqnarray}
1 \ &=& \ c\int_{{\mathcal{S}}^2}d{\bf m} \ |\langle 0({\bf m}')|0({\bf
m})\rangle|^2
\ = \ c \frac{4\pi}{2j + 1}\, .
\end{eqnarray}
So finally the resolution of the unity may be written in one of the
following equivalent forms:
\begin{eqnarray}
\hat{\mathbf{1}} \! = \! \frac{2j + 1}{4\pi}\int_{{\mathcal{S}}^2}\!\!d{\bf
m} \ |0({\bf m})\rangle\langle 0({\bf m})|
\! = \!  \frac{2j + 1}{\pi} \int_{{\mathcal{S}}^2} \frac{d\xi
d\xi^*}{(1+|\xi|^2)^2} \ |0(\xi^*)\rangle \langle 0(\xi) |,
\label{res_un_I}
\end{eqnarray}
where in the last line we have used
\begin{eqnarray*}
d\xi d\xi^* \ \equiv \ d\Real \xi \ d\Imag \xi \, ,
\end{eqnarray*}
with $\Real$ and $\Imag$ denoting the real and  imaginary parts,
respectively.

\section{$SU(2)$ CS functional
integral\label{ApJ}}
\subsection{Generalized coherent
states and functional integrals
\label{AJK}}

We are now in position  to construct the corresponding
functional-integral representation of a transition amplitude $\langle
0({\bol{\zeta}}_f),t_f|0({\bol{\zeta}}_i),t_i\rangle$.
Similarly as in the usual functional-integral constructions~\cite{Kleinert} the key is the
Heisenberg-picture resolution of unity
that in the present case reads (cf Eq.~(\ref{1.2.3.38}))
\begin{eqnarray}
\hat{\mathbf{1} } \ = \ c
\int_{G/H}d{\bol{\zeta}} \ |0({\bol{\zeta}}),t\rangle\langle 0({\bol{\zeta}}),t|\, .
\label{AJK1a}
\end{eqnarray}
The latter holds for all times $t$. Let us now
partition the time interval $[t_i,t_f]$ into $N+1$ equidistant
pieces $\Delta t$
by writing $t_f - t_i =
(N+1)\Delta t$.
%
%
%
We can now label the intermediate times as, say $t_n = t_i + n\Delta t$, $n = 1, 2, \ldots, N$.
Introducing the resolution of unity for every intermediate time point, we obtain
\begin{eqnarray}
\langle 0({\bol{\zeta}}_f),t_f|0({\bol{\zeta}}_i),t_i\rangle \ &=& \
\left( \int_{G/H} \ \! \prod_{k=1}^N c \ d{\bol{\zeta}}_k \right)\ \langle 0({\bol{\zeta}}_f)
,t_f|0({\bol{\zeta}}_{N}),t' -\Delta t \rangle\nonumber \\[2mm]&&
\times \ \langle 0({\bol{\zeta}}_N),t'-\Delta t |0({\bol{\zeta}}_{N-1}),t' -2\Delta t \rangle\nonumber \\[2mm]
&& \times \ \langle 0({\bol{\zeta}}_{N-1}),t'-2\Delta t |0({\bol{\zeta}}_{N-2}),t' -3\Delta t \rangle\nonumber \\
&& \ \mbox{\hspace{2cm}} \vdots \nonumber \\
&& \times \ \langle 0({\bol{\zeta}}_{1}),t+ \Delta t
|0({\bol{\zeta}}_i),t_i \rangle\, .
\end{eqnarray}
We have formally set $t_0 = t_i$ and $t_{N+1} = t_f$. The affiliated
infinitesimal-time transition amplitude can be written as
\begin{eqnarray}
&&\langle 0({\bol{\zeta}}_{k}),t_k |0({\bol{\zeta}}_{k-1}),t_{k-1}\rangle \
\simeq \ \langle 0({\bol{\zeta}}_{k})|\left( 1 - {i} \!
\int_{t_{k-1}}^{t_k}
dt \ \hat{H}(t) \right) |0({\bol{\zeta}}_{k-1})\rangle \nonumber \\[2mm]
&&\mbox{\hspace{20mm}}\simeq \ \langle 0({\bol{\zeta}}_{k})|0({\bol{\zeta}}_{k-1})\rangle \left(1-
i{\Delta t}\ H({\bol{\zeta}}_k, {\bol{\zeta}}_{k-1}, t_k)\right)\nonumber \\[2mm]
&&\mbox{\hspace{20mm}}\simeq \ \langle 0({\bol{\zeta}}_{k})|0({\bol{\zeta}}_{k-1})\rangle \ \exp\left(- {i} \! \int_{t_{k-1}}^{t_k}\!\!dt \
H({\bol{\zeta}}, \dot{{\bol{\zeta}}}, t) \right) \, .
\label{Eq.25a}
\end{eqnarray}
Here
\begin{eqnarray*}
H({\bol{\zeta}}_k, {\bol{\zeta}}_{k-1},t_k) \ = \ \frac{\langle 0({\bol{\zeta}}_{k})|\hat{H}(t_k)|0({\bol{\zeta}}_{k-1})\rangle}{\langle 0({\bol{\zeta}}_{k})|0({\bol{\zeta}}_{k-1})\rangle}\, ,
\end{eqnarray*}
is the normalized matrix element of the Hamiltonian.
Eq.~(\ref{Eq.25a}) can be further simplified if we use the fact that
to the leading order in $\Delta t$
\begin{eqnarray}
\langle 0({\bol{\zeta}}_{k})|0({\bol{\zeta}}_{k-1})\rangle  &\simeq& \ 1 -
\langle 0({\bol{\zeta}}_{k})|\{|0({\bol{\zeta}}_{k})\rangle  - |0({\bol{\zeta}}_{k-1})\rangle \}\nonumber \\[3mm]
&\simeq & \exp\left(  -\Delta t \  \frac{\langle 0({\bol{\zeta}}_{k})|\{|0({\bol{\zeta}}_{k})\rangle  - |0({\bol{\zeta}}_{k-1})\rangle \}
}{\Delta t} \right)\nonumber \\[2mm]
&\simeq& \exp\left(- \int_{t_{k-1}}^{t_k} \langle 0({\bol{\zeta}})|\frac{d}{dt} |0({\bol{\zeta}})\rangle \ dt\right)\, .
\label{Eq.26a}
\end{eqnarray}
It should be also noted that both
$|0({\bol{\zeta}}_j)\rangle$ and $\langle 0({\bol{\zeta}}_i)|$  are now the {\em Schr\"{o}dinger-picture} CS.
Combining Eq.~(\ref{Eq.25a}) together with Eq.~(\ref{Eq.26a}) allows to write the finite-time transition amplitude
in the large $N$ limit as
\begin{eqnarray}
&&\mbox{\hspace{-0mm}}\langle 0({\bol{\zeta}}_f),t_f|0({\bol{\zeta}}_i),t_i\rangle \nonumber \\[2mm]
&&\mbox{\hspace{0mm}}=  \int_{{\tiny{\bol{\zeta}}}(t_i) = {\bol{\zeta}}_i}^{{\bol{\zeta}}(t_f) = {\bol{\zeta}}_f}
{\mathcal{D}}\mu({\bol{\zeta}})\exp\left({i}\!\int_{t_i}^{t_f}\!\!dt
\left[ \langle 0({\bol{\zeta}})|i\frac{d}{dt} |0({\bol{\zeta}})\rangle
 - H({\bol{\zeta}}, \dot{\bol{\zeta}}, t)\right]\right)\!.\label{tran_amp_III}
\end{eqnarray}
Here we have formally identified the functional-integral measure as
\begin{eqnarray}
\int_{{\bol{\zeta}}(t_i) = {\bol{\zeta}}_i}^{{\bol{\zeta}}(t_f) = {\bol{\zeta}}_f}  {\mathcal{D}}\mu({\bol{\zeta}}) \cdots \ = \ \lim_{N\rightarrow\infty}\left( \int_{G/H} \ \!
\prod_{k=1}^N c \ d{\bol{\zeta}}_k \right) \cdots\, .
\end{eqnarray}
Let us also observe that the assumed {\em square
integrability} of generalized CS implies
\begin{eqnarray}
\langle 0({\bol{\zeta}})|i\frac{d}{dt} |0({\bol{\zeta}})\rangle  \ = \
-\frac{d}{dt}\{\langle 0({\bol{\zeta}})|\} i |0({\bol{\zeta}})\rangle \ = \
(\langle 0({\bol{\zeta}})|i \frac{d}{dt} |0({\bol{\zeta}})\rangle)^*\, ,
\label{Eq.28a}
\end{eqnarray}
i.e., $\langle 0({\bol{\zeta}})|i d/dt |0({\bol{\zeta}})\rangle$ is {\em purely
real}. There is an intimate connection of (\ref{Eq.28a}) with the concept of {\em geometric phase}.
To see this we write the corresponding phase factor appearing in the path integral
(\ref{tran_amp_III}) as
\begin{eqnarray}
\int_{t_i}^{t_f}  \langle 0({\bol{\zeta}})|i \frac{d}{dt} |0({\bol{\zeta}})\rangle \ dt \ = \
\int_{\gamma}
 \langle 0({\bol{\zeta}})| i \vect{\nabla}_{\!\tiny{\mbox{${\bol{\zeta}}$}}}
| 0({\bol{\zeta}})\rangle\cdot d{\bol{\zeta}}\, . \label{K.25aac}
\end{eqnarray}
In particular, when $|0({\bol{\zeta}})\rangle$ are eigenstates of the
Hamiltonian (as, for instance, in non-linear $\sigma$ models
where $|0({\bol{\zeta}})\rangle$ describe the degenerate ground-state) and
when  ${\bol{\zeta}}(t)$  traverses during the period $t_f- t_i$ a closed
path $\gamma$ in the $G/H$ space, then
(\ref{K.25aac}) corresponds to the fundamental formula
for the Berry--Anandan phase~\cite{Berry:,Anandan,BJ}.
Closed paths typically occur when (quantum-mechanical) partition functions $Z$ are to be computed~\cite{Kleinert}. This is because in such a case
\begin{eqnarray}
\int_{{\bol{\zeta}}(t_i) = {\bol{\zeta}}_i}^{{\bol{\zeta}}(t_f) = {\bol{\zeta}}_f} {\mathcal{D}}\mu({\bol{\zeta}}) \cdots \ \mapsto \ \int_{G/H} d{\bol{\zeta}}_i \int_{{\bol{\zeta}}(t_i) = {\bol{\zeta}}_i}^{{\bol{\zeta}}(t_f) = {\bol{\zeta}}_i} {\mathcal{D}}\mu({\bol{\zeta}}) \cdots \, .
\end{eqnarray}
We shall say more on this in
Section~\ref{ApK3}.

\subsection{$SU(2)$ coherent states\label{SU(2)cs}}

Results of the previous two subsections can be now particularized for
the $SU(2)$ CS. Namely, from Eq.~(\ref{tran_amp_III}) the
transition amplitude can be written in the form
\begin{eqnarray}
&&\mbox{\hspace{-2mm}}\langle 0(\xi^*_f),t_f| 0(\xi_i),t_i\rangle
\ = \ \lim_{N\rightarrow\infty} \left(\int\prod_{k=1}^{N} d\mu(\xi^*_k,\xi_k)\right)\nonumber \\[2mm]
&&\mbox{\hspace{25.5mm}}\times \ \exp\left({i} \sum_{l=0}^N \Delta
t\!\left[ \frac{i}{\Delta t} \langle
0(\xi^*_l)|\Delta| 0(\xi_{l})\rangle - H(\xi_l^*,\xi_{l-1},t_l)
\right]   \right) \nonumber \\[2mm]
&&\mbox{\hspace{-2mm}}= \ \int_{\xi(t_i) = \xi_i}^{\xi^*(t_f) =
\xi^*_f} {\mathcal{D}}\mu(\xi^*,\xi) \exp\left({i}
\!\int_{t_i}^{t_f}\!dt\left[ \langle
0(\xi^*)|i\frac{d}{dt}|0(\xi)\rangle - H(\xi^*,\xi, t)\right]
\right)\nonumber \\[2mm]
&&\mbox{\hspace{-2mm}}= \ \int_{\xi(t_i) = \xi_i}^{\xi^*(t_f) =
\xi^*_f} {\mathcal{D}}\mu(\xi^*,\xi) \exp\left({i}\!\!
\int_{t_i}^{t_f}\!dt\left[i\ \!\frac{j(\xi^*\dot{\xi} -
\dot{\xi^*}\xi )}{(1+|\xi|^2)} - H(\xi^*,\xi, t) \right]
\right). \label{appK.25aa}
\end{eqnarray}
Here
\begin{eqnarray*}
\mbox{\hspace{-3mm}}d\mu(\xi^*_k,\xi_k) \ \equiv \ \frac{d\xi_k
d\xi^*_k}{(1+|\xi_k|^2)^2}\;\;\ \mbox{and} \;\;\
H(\xi_l^*,\xi_{l-1},t_l) \ \equiv \ \frac{\langle 0(\xi^*_{l})|
{H}(t_l)|0(\xi_{l-1})\rangle}{\langle 0(\xi_l^*)| 0(\xi_{l-1})
\rangle}\, .
\end{eqnarray*}
Use was also made of the fact that up to the order $\Delta \xi_l =
\xi_l - \xi_{l-1}$ one has
\begin{eqnarray*}
\mbox{\hspace{-3mm}}\langle 0(\xi^*_l)|\Delta| 0(\xi_{l})\rangle \ = \ \langle
0(\xi^*_l)|\{ | 0(\xi_{l})\rangle - | 0(\xi_{l-1})\rangle
\}\nonumber  \ = \ \frac{j(\xi_l^*\Delta \xi_l - \xi_l \Delta
\xi_l^*)}{1+ |\xi_l|^2} \, .
\end{eqnarray*}
%

Path integral for $SU(2)$ CS was originally
constructed by Klauder~\cite{Klauder:1979gi} and Kuratsuji and
Suzuki~\cite{Kuratsuji:1980kf}. Its main utility has been in
semiclassical treatments of quantum systems which have Hamiltonians
composed of the generators of the $SU(2)$ group,
although other applications,
such as duality or geometrical phases of spin systems, are also
frequently mentioned in the literature.

Generalization to field theory (e.g., to continuous spin lattice) can now proceed along standard
lines. In particular one formally
exchanges the coset-space variables $\zeta^a(t)$ ($a = 1, \ldots ,
\dim{G/H}$) with coset-space fields
$\phi^a({\bf x}, t)$. These fields provide a mapping from
$D\!+1$-dimensional spacetime to group quotient $
G/H$, i.e.,
%
$\phi^a({\bf x}, t): \ \mathbb{R}^{D+1}  \mapsto \ G/H$.
%
The space $G/H$, into which the mapping is done, is known
as the {\em target space}.

\section{Nambu--Goldstone theorem and the structure of vacuum manifold\label{ApJ0}}

We begin this section by summarizing the quantum field theory procedure leading
to the Nambu--Goldstone theorem~\cite{BJV,Goldstone}. This is of course well known
but it is useful to repeat it here in order to make our discussion
self-contained. We will also need it in Section~\ref{ApK3} in order
to set up functional integrals for NG fields
and to correctly interpret the ensuing results.
Briefly stated, the theorem asserts
that for a physical system with a {\em global internal}
symmetry group $G$ which is spontaneously broken down
to a subgroup $H$, there are $\mbox{dim}(G/H) = \mbox{dim}G - \mbox{dim}H$  massless modes --- NG bosons.
For our purpose the best way to introduce the NG theorem is
to use the Lorentz-invariant setting and apply the coset space
construction of SSB~\cite{BJV}.
A non-relativistic variant of the theorem will be discussed subsequently.

Let us assume that a full symmetry group of the system, the so-called {\em disordered-phase} symmetry,
is $G$. The
Hamiltonian is thus invariant under action of $G$, i.e.
\begin{eqnarray}
\hat{D}^{-1}(g)\hat{H}\hat{D}(g)  \ = \ \hat{H} \;\;\;\; \mbox{for} \;\;\;\;
\forall g\in G\, . \label{7.1.117}
\end{eqnarray}
Here $\hat{D}(g)$ is a {\em unitary} operator representing the element $g\in
G$ in Hilbert space.  The SSB occurs when the vacuum is invariant {\em only}
under some subgroup
$H$ of $G$. This, for instance, happens when the system is cooled down
below a critical temperature $T_{c}$.
A hallmark of the SSB is the existence of some operator
$\hat{\Phi}$ known as the {\em order
parameter}~\cite{Landau:1996a} whose ground-state
expectation value $\Phi^0$  is not invariant under the whole group $G$, but
only under  $H$.
The symmetry $H$ is
known as the {\em broken-phase} or {\em ordered-phase} symmetry.

Let us for definiteness consider the order parameter to be a
multiplet  ${\bf{\hat{\Phi}}}$
transforming under some $n$-dimensional
representation $S$ of $G$, i.e.
\begin{eqnarray}
\hat{D}^{-1}(g)\hat{\Phi}_i\hat{D}(g) \ = \ \sum_{j =1}^n S_{ij}(g)
\hat{\Phi}_j\, .
\end{eqnarray}
By definition, the vacuum expectation value $\langle 0|
\hat{\Phi}_i|0\rangle \equiv {\Phi}_i^0$ is not invariant under
whole $G$ but only under $H$. This means that for $g$ from $G/H$
\begin{eqnarray}
\langle 0|\hat{D}^{-1}(g) \hat{\Phi}_i\hat{D}(g)|0\rangle \ = \ \sum_{j
=1}^n S_{ij}(g){\Phi}_j^0 \ \neq \ {\Phi}_i^0\, . \label{7.1.119}
\end{eqnarray}
On the level of group generators this may be phrased as
\begin{eqnarray}
\sum_{j
=1}^n S_{ij}(T^a){\Phi}_j^0 \ \neq \ 0  \,\,\,\,\,\mbox{and}\,\,\,\,\, \sum_{j
=1}^n S_{ij}({t^r}){\Phi}_j^0 \ = \ 0\, ,
\label{7.1.120a}
\end{eqnarray}
where ${t^r}$ are generators from $H$ and $T^a$ are broken-symmetry generators.
Eq.~(\ref{7.1.119}) clearly shows that the ground state is not invariant under
the action of $g \in G/H$, i.e.,
\begin{eqnarray}
\hat{D}(g)|0\rangle \ \equiv \ |0(g) \rangle \ \neq \ |0\rangle \;\;\;\;
\mbox{for} \;\;\;\;  g \in G/H \, ,
\end{eqnarray}
or equivalently $\hat{D}(T^a)|0\rangle \neq  0$. Since states $|0(g)\rangle $ are also
eigenstates of $\hat{H}$ with the same eigenvalue as $|0\rangle $ (cf. Eq.~(\ref{7.1.117})), the ground state
is degenerate and distinct states are distinguished
by different $g$'s from $G/H$. So the manifold of degenerate vacuum states
--- {\em vacuum manifold},
can be identified with the quotient space $G/H$.

To proceed we note that (\ref{7.1.119}) can be around a unit element written for all ``$a$"  as
%
\begin{eqnarray}
\lim_{V\rightarrow \infty}\langle 0|[\hat{Q}_V^a(t), \hat{\Phi}_i(0)]|0\rangle
\ = \ \sum_{j=1}^n {S}_{ij}(T^a){\Phi}_j^0 \ \neq \ 0\, .
\label{3.28b}
\end{eqnarray}
Here $\hat{Q}_V^a(t)$ is the regularized Noether charge associated with the generator $T^a$, namely
\begin{eqnarray}
\hat{Q}_V^a(t) \ = \ \int_V d{\bf x} \ \! \hat{J}_0^a({\bf x},t) \, ,
\end{eqnarray}
where $\hat{J}_0^a({\bf x},t)$ is the conserved Noether current.
In (\ref{3.28b}) we have used the translational invariance of the vacuum which allowed us to work with $\hat{\Phi}_i(0)$. The regularization used in Eq.~(\ref{3.28b}) is necessary since $\hat{Q}^a$ is not mathematically well defined --- it is not {\em unitarily implementable}~\cite{BJV}. Indeed, the translation invariance of the vacuum implies that
\begin{eqnarray}
\langle 0| \hat{Q}^{\ \!\!a} \hat{Q}^{\ \!\!a}|0 \rangle \ = \ \int d{\bf x} \ \! \langle 0|\hat{J}_0^a({\bf x},t) \hat{Q}^{\ \!\!a}|0 \rangle\, ,
\end{eqnarray}
is divergent. Inserting now a complete set of intermediate energy states and using again the  translational invariance of the vacuum we get from (\ref{3.28b})
\begin{eqnarray}
&&\lim_{V\rightarrow \infty}\sum_n \int_V d{\bf x} \left[ \langle 0| \hat{J}_0^a (0)|n\rangle\langle n|\hat{\Phi}_i(0) |0\rangle \ \! e^{-i x p_n} \right.\nonumber \\[0mm]
&&\mbox{\hspace{2.4cm}}-\left. \langle 0| \hat{\Phi}_i(0)|n\rangle\langle n| \hat{J}_0^a (0)|0\rangle \ \! e^{i x p_n}\right]\nonumber \\[2mm]
&&= \sum_n (2\pi)^D \delta^{(D)}({\bf p}_n)\left[ \langle 0| \hat{J}_0^a (0)|n\rangle\langle n|\hat{\Phi}_i(0) |0\rangle \ \! e^{-i E_n t} \right.\nonumber \\[0mm]
&&\mbox{\hspace{2.4cm}}-\left. \langle 0| \hat{\Phi}_i(0)|n\rangle\langle n| \hat{J}_0^a (0)|0\rangle \ \! e^{i E_n t}\right] \ \neq \ 0\, .
\label{3.29a}
\end{eqnarray}
Here $p_n =(E_n, {\bf p}_n)$ and $D$ is the spatial dimension.
As long as the theory satisfies the {\em microcausality} condition, i.e., the commutator of any two local operators separated by a space-like interval vanishes, we have
\begin{eqnarray}
\frac{d}{d t}[\hat{Q}_V^a(t), \hat{\Phi}_i(0)] \ &=& \  \int_V d{\bf x} \  \![\partial^{\mu} \hat{J}_{\mu}^{a}({\bf x},t), \hat{\Phi}_i(0)]\nonumber \\ &-& \ \oint_{\Sigma} dS^i [\hat{J}^a_i({\bf x},t), \hat{\Phi}_i(0)] \ \stackrel{V\rightarrow \infty}{\longrightarrow} \ 0\, .
\end{eqnarray}
%
$\Sigma$ denotes the surface bounding the volume $V$, i.e. the sphere $S^{D-1}$.
This indicates that after the time derivative the last two lines of (\ref{3.29a}) give
\begin{eqnarray}
&&\sum_n (2\pi)^D \delta^{(D)}({\bf p}_n)E_n \left[ \langle 0| \hat{J}_0^a (0)|n\rangle\langle n|\hat{\Phi}_i(0) |0\rangle \ \! e^{-i E_n t} \right.\nonumber \\[0mm]
&&\mbox{\hspace{3.4cm}}+ \left. \langle 0| \hat{\Phi}_i(0)|n\rangle\langle n| \hat{J}_0^a (0)|0\rangle \ \! e^{i E_n t}\right] \ = \ 0\, .
\label{3.32a}
\end{eqnarray}
Comparing (\ref{3.29a}) with (\ref{3.32a}) shows that there must exist a state $|n\rangle$ such that
\begin{eqnarray}
\langle 0| \hat{\Phi}_i(0)|n\rangle\langle n| \hat{J}_0^a (0)|0\rangle \ \neq \ 0 \;\;\;\;\mbox{for}  \;\;\;\; \delta^{(D)}({\bf p}_n)E_n \ = \ 0\, .
\label{4.44}
\end{eqnarray}
This state is a massless state with the same quantum number as $\hat{Q}^{\ \!\!a}$  since it is generated by
$\hat{Q}^{\ \!\!a}$ from the vacuum $|0\rangle$.
In particular, the field excitations corresponding to this state (the so-called
the NG excitations)
must have the same Lorentz properties as the charge $\hat{Q}^{\ \!\!a}$. Because the charge is related to internal symmetries,  the NG field must be a Lorentz scalar (or pseudo-scalar) and a boson. A similar argument for
spontaneously-broken {\em supersymmetry} implies that the NG
particles are spin-$1/2$ fermions, and they are spin-$1$ bosons
(e.g., phonons) for spontaneously-broken {\em translation} invariance.

Let us define the vacuum state
%
$|0({\bol{\pi}})\rangle   \ \equiv \   \exp(i{\bol{\pi}}\!\cdot\hat{Q})|0\rangle$
%
where ${\bol{\pi}}\cdot\hat{Q} = \pi_{a} \hat{Q}^a$.
If we consider in the neighborhood of  the vacuum state $|0({\bol{\pi}})\rangle$
an infinitesimal transformation ${\bol{\theta}}$, say in the direction ``$a$", we obtain (no summation over ``$a$")
\begin{eqnarray}
\delta_\theta |0({\bol{\pi}})\rangle  \ &=& \
\exp(i \theta_a \hat{Q}^a)|0({\bol{\pi}})\rangle - |0({\bol{\pi}})\rangle
\ = \ i \theta_a \hat{Q}^a |0({\bol{\pi}})\rangle\, .
\label{4.45a}
\end{eqnarray}
%
Because the argument leading to (\ref{4.44}) could be repeated for any ground state $|0(g)\rangle$, $g\in G/H$,
Eq.~(\ref{4.45a}) implies that $\delta |0({\bol{\pi}})\rangle \propto | n \rangle$ for any ${\bol{\pi}}$. So
the NG state corresponds to a shift within the vacuum manifold (shift along ``flat energy directions").
In this respect, the NG fields give a meaning to the fluctuations among degenerate ground states.
Note that the field which $\delta_{\theta}$-fluctuates in $a$-th energy flat direction can be associated with the group parameter $\theta_a$.
One may thus identify the local group parameters ${\bol{\theta}}$ with the NG multiplet.
Since at every point ${\bol{\pi}}$ of the vacuum manifold there are $\mbox{dim}(G/H)$ independent flat directions (namely independent tangent directions of the {\em local frame} in ${\bol{\pi}}$), there must be  $\mbox{dim}(G/H)$  distinct NG fields.
So ${\bol{\theta}}$ form a local coordinate system at ${\bol{\pi}}$.
Starting with a fixed ${\bol{\pi}}$, one may extend the local domain of ${\bol{\theta}}$ globally on the whole $G/H$ by applying the transformation rules for broken symmetries in $G/H$ on the parameters ${\bol{\theta}}$. The involved mathematical technicalities are most easily done through the  {\em Maurer--Cartan one-forms}~\cite{Burgess}.  Extension of the NG fields  on the whole $G/H$ allows to put in one-to-one connection
the NG fields and points on  $G/H$.
In this way,
the NG fields {\em coordinatize}
the quotient space $G/H$.

Alternatively, one may view the NG modes as representing the fluctuations in the order parameter. Indeed, using (for a simplicity of the argument) the vacuum state at ${\bol{\pi}} = 0$ we can write (no summation over ``$a$")
%
\begin{eqnarray}
\lim_{V\rightarrow\infty}\langle 0|i\theta_a \ \![\hat{Q}_V^a(t), \hat{\Phi}_i(0)]|0\rangle
\ = \
\delta{\Phi}_i^0 \ = \
i\theta_a S_{ij}(T^a){\Phi}_j^0\, .
\label{4.46.a}
\end{eqnarray}
%
%
%
From our previous discussion follows that the local parameter $\theta_a$  coincides with the near-to-origin NG field, and so
the $\delta{\Phi}_i^0$ is directly proportional to the NG field.
The preceding equation is often a reason why some people normalize the NG field in such a way that  $\theta_a S_{ij}(T^a){\Phi}_j^0$  itself is considered as the definition of the NG field~\cite{Burgess}.

As shown in Section~\ref{ApD}, the group quotient $G/H$
can be identified with a set of all generalized
CS corresponding to the group $G$. Connection with a vacuum manifold is then
established when as a fiducial vector one chooses any ground-state vector $|0(g)\rangle$.

Let us finally stress that the NG theorem is valid, with few qualifications, even for non-Lorentz-invariant situations such as those that occur frequently in solid-state physics.  The caveat in the above proof is the use of translational invariance and microcausality.  In particular the microcausality should be in the non-relativistic setting substituted with an absence of long-range interactions~\cite{Nielsen:1976}. Under assumption that the translational invariance is not broken it can be showed that the total number of NG bosons might be less than the number of broken generators, in contrast to the naive expectation based on experience with Lorentz-invariant systems. The precise rule for counting the NG modes can be found, e.g., in Ref.~\cite{Nielsen:1976}.

Fortunately the NG fields serve also in the non-relativistic framework as coordinates
on the vacuum manifold $G/H$.
The point is that the number of NG fields still coincide with the number of broken generators, it is only that the number of NG fields does not match the number of NG bosons. Connection between broken generators and NG bosons depends in  a non-relativistic context on the dispersion relation. This will be explicitly illustrated in the following section.
%



\section{$SU(2)/U(1)\!-\!\sigma$ model and Landau--Lifshitz ferromagnetic magnons\label{ApK3}}


Because the functional integrals based on generalized coherent states are naturally phrased
in terms of coset-space variables they are well suited to describe the effective low-energy dynamics
of theories with SSB.
In particular, when $G$ is the disordered-phase symmetry and $H$ is
the broken-phase symmetry then the NG fields take
values in the target space which is a coset of groups $G/H$. More details can be found, e.g., in
Ref.~\cite{BJV}. Massless field theories where the target space is the group coset
space $G/H$ are commonly known as $G/H$-$\sigma$ models or also
as non-linear $\sigma$ models.
With a suitable choice of the Hamiltonian $H({\bol{\zeta}}, \dot{{\bol{\zeta}}},
t)$ will the generalized CS functional integrals (and the associated non-linear $\sigma$ models)
describe low-energy effective field
theories, in which only NG bosons,
including their mutual interactions, will propagate.

NG bosons are true dynamical protagonists in many
low-energy or low-temperature solid-state systems. In this respect it
is instructive to consider some representative system where one can
explicitly see how the correct NG dynamics is reproduced via generalized
CS path integrals. Along these lines we now derive
the correct behavior of ferromagnetic
magnons  in the Heisenberg model of ferromagnets. This problem
was historically notoriously difficult to deal with. In particular,
the usual mean-field approaches fail to provide the
quadratic dispersion behavior which is typically observed
in inelastic scattering of spin-polarized neutrons by magnons.
Since ferromagnetic materials are paradigmatic examples of
systems with SSB~\cite{BJV} --- the disordered-phase symmetry $SU(2)$ is below the Curie temperature
spontaneously broken to the residual rotational symmetries $U(1)$ --- it is only natural
to use the $SU(2)/U(1)\!-\!\sigma$ model to deal with the corresponding low-energy degrees of freedom.
The resulting gapless NG modes should be then identifiable with scalar bosonic excitations
around the ground state of the spin-$j$ Heisenberg
ferromagnets. The only experimentally viable candidates for such excitations are the gapless spin waves known as magnons.
%
%
%
%
%
By following this reasoning we show that in the long-wavelength limit
one can obtain the Landau--Lifshitz non-linear $\sigma$ model which describes the correct
dynamics and dispersion relations for ferromagnetic magnons.

To see how all this comes about we first rewrite the action in the
path integral (\ref{appK.25aa}) in terms of the unit-vector
dynamical variables ${\bf n}(t)$. The first term can be then expressed as
\begin{eqnarray}
i\frac{j(\xi^*d{\xi} - d{\xi^*}\xi )}{(1+|\xi|^2)} \ &=& \ -2j
\sin^2(\theta/2)\ \! d{\varphi} \ = \ - \frac{j}{r(z+r)}\ \!(x\ \!
dy - y\ \! dx) \nonumber \\[1mm]
&=& \ {\bf{A}}_B({\bf x}) \cdot d{\bf x}\, ,
\end{eqnarray}
where the vector potential
\begin{eqnarray}
{\bf{A}}_B({\bf x}) \ = \ - \frac{j}{r(z+r)}\ \! (-y,x,0)\, ,
\end{eqnarray}
corresponds to Berry's connection. Since the vector ${\bf x}$  sweeps the surface of
${\mathcal{S}}^2$ we have that ${\bf x} = {\bf n}$ (${\bf n}^2 =
1$). The first term in the action in (\ref{appK.25aa}) thus reads
\begin{eqnarray}
i\int_{t_i}^{t_f} \! \!dt \ \! \frac{j(\xi^*\dot{\xi} -
\dot{\xi^*}\xi )}{(1+|\xi|^2)} \ = \ \int_{t_i}^{t_f} \!
{\bf{A}}_B({\bf n})\cdot \frac{d{\bf n}}{d t} \ dt \ = \
\int_{\Sigma} {\bf B}_B \cdot d {\vect \sigma}\, .\label{K.31ab}
\end{eqnarray}
With $\Sigma$ denoting area of ${\cal S}^2$ bounded by a closed
loop traversed by ${\bf n}(t)$. Berry's magnetic induction
${\bf B}_B$ has the explicit form
\begin{eqnarray}
{\bf B}_B({\bf x}) \ = \ \vect{\nabla} \wedge {\bf{A}}_B({\bf x})\ =
\  \frac{j}{r^3}\ \! {\bf x} \ = \ \frac{j}{r^2}\ \! {\bf n} \ = \ j
{\bf n} \, , \label{K.32aa}
\end{eqnarray}
which implies that
\begin{eqnarray}
\int_{{\mathcal{S}}^2} {\bf B}_B \cdot d {\vect \sigma} \ = \ 4\pi
j\, . \label{K.33aa}
\end{eqnarray}
Eq.~(\ref{K.32aa}) together with (\ref{K.33aa}) shows that there is a
monopole of the magnetic charge $j$ located in the origin of our
target space. We also note that from (\ref{K.31ab}) and
(\ref{K.32aa}) follows
%
\begin{eqnarray}
i\!\int_{t_i}^{t_f} \! \!dt \ \! \frac{(\xi^*\dot{\xi} -
\dot{\xi^*}\xi )}{(1+|\xi|^2)} \ &=& \ \int_0^1\! du
\int_{t_i}^{t_f}\!\! dt \ {\bf n}(t,u)\cdot [\partial_t{\bf
n}(t,u) \wedge
\partial_{u}{\bf n}(t,u)]\nonumber \\[1mm]
&\equiv& \  S_{WZ}[{\bf n}] \, , \label{K.34ab}
\end{eqnarray}
where ${\bf n}(t,u)$ is an arbitrary extension  of ${\bf n}(t)$ into
the spherical rectangle defined by the limits of integration and
fulfilling conditions: ${\bf n}(t,0) = {\bf n}(t)$, ${\bf n}(t,1) =
(1,0,0)$, and ${\bf n}(t_i,u) = {\bf n}(t_f,u)$. The $S_{WZ}[{\bf
n}]$ is a special member  of a wide class of actions known as
Wess--Zumino actions
~\cite{Witten:1979b}. Eq.~(\ref{K.34ab}) then demonstrates a
typical situation ubiquitous in effective theories, namely, that Berry--Anandan
phase gives rise to the Wess--Zumino action. Examples include low-dimensional ferromagnets
with local anisotropies~\cite{Braun} or non-abelian gauge theories with topological angle ($\theta$-term)~\cite{Simons}.

Let us now turn to many-spin systems and consider a lattice of spins. We will
concentrate first on the Hamiltonian $H(\xi^*,\xi,t)$. To this end
we consider the Hamiltonian for the ferromagnetic Heisenberg model,
i.e.
\begin{eqnarray}
\hat{H}({\bf J}) \ = \ K \sum_{\{{\bf x}, {\bf x}'\} } \hat{\bf
J}({\bf x})\cdot \hat{\bf J}({\bf x}')\, .
\end{eqnarray}
where $K = -|K|$ is the Heisenberg exchange constant and $\{{\bf x}, {\bf x}'\}$
denotes pairs of neighboring lattice sites. According to the definition
of $H(\xi^*_k,\xi_{k-1},t)$ we have
\begin{eqnarray}
H(\xi^*_k,\xi_{k-1},t) \ &=& \ H({\bf n}_k, {\bf n}_{k-1}) \ = \
\frac{\langle 0({\bf n}_k)| \hat{H}({\bf J})| 0({\bf
n}_{k-1})\rangle}{\langle 0({\bf n}_k)|({\bf
n}_{k-1})\rangle}\nonumber \\[1mm]
&\approx& \ \langle 0({\bf n}_k)| \hat{H}({\bf J})| 0({\bf
n}_{k})\rangle \ + \ \mathcal{O}(\Delta  t)\, .
\end{eqnarray}
By taking advantage of the identity  $\langle 0({\bf n}_k)|\hat{\bf
J}({\bf x})| 0({\bf n}_{k})\rangle =  j {\bf n}_k({\bf x})$
we obtain
\begin{eqnarray}
\ H({\bf n}_k, {\bf n}_{k-1}) \ \approx \ -|K| j^2\sum_{\{{\bf x}, {\bf x}'\}}
{\bf n}_k({\bf x})\cdot {\bf n}_k({\bf x}')\, ,
\end{eqnarray}
so that action in the functional integral (\ref{appK.25aa}) reads
\begin{eqnarray}
\mbox{\hspace{-0mm}}S[{\bf n}] \ = \    j \sum_{{\bf x}} S_{WZ}[{\bf
n}({\bf x})] \ + \ |K| j^2 \sum_k \Delta t\sum_{\{{\bf x}, {\bf x}'\}
} {\bf n}_k({\bf x})\cdot {\bf n}_k({\bf x}')\, .
\end{eqnarray}
Here the first sum runs over all the sides of the lattice and thus
represents the sum of the Wess--Zumino terms of individual spins.
Note particularly, that the time derivative (and hence dynamics)
enters only through the Wess--Zumino term.

For definiteness sake we now consider a $D$-dimensional hypercubic
lattice and restrict $\sum_{\{{\bf x}, {\bf x}'\}}$ to nearest
neighbors only. With this we can write
\begin{eqnarray}
\mbox{\hspace{-0mm}}\sum_{\{{\bf x}, {\bf x}'\} } {\bf
n}_k({\bf x})\cdot {\bf n}_k({\bf x}') \ = \ -
\frac{1}{2}\sum_{\{{\bf x}, {\bf x}'\} }[{\bf n}_k({\bf x}) -
{\bf n}_k({\bf x}')]^2 \ + \ \mbox{const.}\, . \label{K39aa}
\end{eqnarray}
Consider now the long-wavelength limit, in which ${\bf n}_k({\bf
x})$ are smooth functions of ${\bf x}$.
By denoting the lattice
spacing $a$ and taking the $N\rightarrow \infty$ (i.e.,
continuous-time) limit we obtain an effective field theory
described by the action
\begin{eqnarray}
S[{\bf n}] \ &=& \  \frac{j}{a^D} \int_{{\mathbb R}^D} d^D {\bf x}\
\!
S_{WZ}[{\bf n}({\bf x})] \nonumber \\[2mm]
&-& \ \frac{ j^2 |K|}{2 a^{D-2}} \int_{t_i}^{t_f} \!d t
\int_{{\mathbb R}^D}\!\! d^D {\bf x}\  \partial_i {\bf n}({\bf
x},t)\cdot \partial_i {\bf n}({\bf x},t)\, . \label{K40aa}
\end{eqnarray}
In this expression we have dropped the constant term from
(\ref{K39aa}) which is irrelevant for dynamical equations.
In order to deal with the non-trivial measure
${\mathcal{D}}\mu({\bf n})$ in the functional integral, we can
rewrite it as ${\mathcal{D}}\mu({\bf n})\delta[{\bf n}^2 -1]$ where
the integration variables  ${\bf n}$  are not any more restricted to
a target space ${\mathcal{S}}^2$. The functional $\delta$-function
can be elevated into the action via functional Fourier transform
\begin{eqnarray}
\mbox{\hspace{-0mm}}\delta[{\bf n}^2 -1] \ &=& \ \lim_{N\rightarrow
\infty}\prod_{i=1}^N
\delta({\bf n}^2({\bf x}_i,t_i) -1) \nonumber \\
&=& \ \int {\mathcal{D}}\lambda \ \! \exp\left(i
\!\!\int_{t_i}^{t_f} \!d t \int_{{\mathbb R}^D}\!\! d^D {\bf x}\
\!\lambda({\bf x},t) ({\bf n}^2({\bf x},t) -1)\right)\! .
\end{eqnarray}
The latter leads to a new {\em total} action
\begin{eqnarray}
S_{\rm tot}[{\bf n}] \ &=& \ S[{\bf n}] \ + \ \int_{t_i}^{t_f} \!d t
\int_{{\mathbb R}^D}\!\! d^D {\bf x}\ \!\lambda({\bf x},t) ({\bf
n}^2({\bf x},t) -1)\, . \label{K42aa}
\end{eqnarray}

Let us now look at the classical equation of motion whose solution
should represent the dominant field configuration in a semiclassical  WKB
approach to quantum ferromagnetism.  The variation $\delta S_{\rm
tot}[{\bf n}] = 0$ implies three equations
\begin{eqnarray}
\mbox{\hspace{-0mm}}j \ \! ({\bf n} \wedge \partial_t {\bf n }) \ +
\ 2 a^D \lambda {\bf n} \ &=& \ - a^2 |K| j^2 \ \! {\vect{\nabla}}^2
{\bf n} \ \ \ \ \ \mbox{and}\ \ \ \ {\bf n}^2 \ = \ 1\, .
 \label{K43aa}
\end{eqnarray}
Here we have employed  that
\begin{eqnarray}
\mbox{\hspace{-0mm}}\delta S_{WZ}[{\bf n}({\bf x})] \ &=& \
\int_0^1\! du \int_{t_i}^{t_f}\!\! dt \ \! \partial_{u}\!\left\{
\delta {\bf n}({\bf x}, t,u)\cdot [{\bf n}({\bf x}, t,u) \wedge
\partial_t{\bf n}({\bf x}, t,u)] \right\}\nonumber \\
&+& \ 3 \int_0^1\! du \int_{t_i}^{t_f}\!\! dt \ \! \delta {\bf
n}({\bf x}, t,u) \cdot [\partial_t{\bf n}({\bf x}, t,\tau) \wedge
\partial_{u}{\bf n}({\bf x}, t,u)]\nonumber \\
&=& \ \int_{t_i}^{t_f}\!\! dt \  \delta {\bf n}({\bf x}, t)\cdot [{\bf n}({\bf x}, t)
\wedge
\partial_t{\bf n}({\bf x}, t)]\, ,
\end{eqnarray}
where the term on the second line is zero because $\partial_t{\bf n} \wedge
\partial_{u}{\bf n}$ is parallel to ${\bf n}$ and  ${\bf n}\cdot\delta
{\bf n} = \delta {\bf n}^2/2 = 0$. On the last line we have used that ${\bf n}(t,0) = {\bf n}(t)$, ${\bf n}(t,1) =
(1,0,0)$. Employing now the
identity ${\bf n}\cdot ({\bf n} \wedge
\partial_t {\bf n }) = 0$ we find for the Lagrange multiplier $\lambda$
\begin{eqnarray}
\lambda \ = \ - \frac{|K|j^2}{2 a^{D-2}} \ { {\bf n} \cdot
{\vect{\nabla}}^2 {\bf n}}\, .
\end{eqnarray}
By inserting this result back into Eq.~(\ref{K43aa}) and applying the identity ${\bf a }\wedge({\bf b }\wedge{\bf c }) =
({\bf a }\cdot {\bf c }){\bf b}  - ({\bf a }\cdot {\bf b }){\bf c }$ we obtain
\begin{eqnarray}
{\bf n }\wedge\left[\partial_t {\bf n } \ - \  a^2 |K| j \ \! ({\bf n } \wedge
{\vect{\nabla}}^2 {\bf n})\right] \ = \ 0 \, .  \label{K46aa}
\end{eqnarray}
Note that both terms inside $[\ldots]$ are orthogonal to ${\bf n }$ and so we can cast the previous equation into a simpler (but equivalent) form, namely
\begin{eqnarray}
\partial_t {\bf n } \ = \  a^2 |K| j \ \! ({\bf n } \wedge
{\vect{\nabla}}^2 {\bf n})  \, .  \label{K46ab}
\end{eqnarray}
%
%
Eq.~(\ref{K46ab}) is known as  Landau--Lifshitz
equation for quantum
ferromagnet~\cite{Landau:1996a}. It
essentially describes the dynamics of a ferromagnetic spin wave. To
see a leading dispersion behavior, we go to the linear regime and
assume that the spins are align around a $3$-rd axis around which
they wobble, or precess, so in particular $n_3$ will change with
$t$ and ${\bf x}$ much slower that $n_{1,2}$. By defining, ${\bf n}
= (\pi_1,\pi_2, \sigma)$ (${\vect \pi}^2 + \sigma^2 =1$), omitting
derivatives of $\sigma$ and setting $\sigma \approx 1$ we linearize
the Landau--Lifshitz equations as
\begin{eqnarray}
\partial_t \pi_1 \ \approx \ - a^2 |K|j  {\vect{\nabla}}^2 \pi_2       \ \ \ \mbox{and}
 \ \ \  \partial_t \pi_2 \ \approx \  a^2 |K|j  {\vect{\nabla}}^2 \pi_1\,
 . \label{K47ab}
\end{eqnarray}
Fourier transform of (\ref{K47ab}) yields the dispersion relation
$\omega({\bf k}) \propto {\bf k}^2$. The modes that obey such a
behavior are {\em ferromagnetic
magnons}. These are true
(non-relativistic) Nambu--Goldstone bosons.
However, notice that the fields $\pi_1$ and $\pi_2$ describe only one NG mode.
This can be understood by rewriting (\ref{K47ab}) equivalently as
\begin{eqnarray}
\partial_t \pi \ \approx \ i a^2 |K|j  {\vect{\nabla}}^2 \pi
      \ \ \ \mbox{and}
 \ \ \  \partial_t \pi^\dag \ \approx \  -i a^2 |K|j  {\vect{\nabla}}^2 \pi^\dag\, ,
 \label{K47abc}
\end{eqnarray}
with $\pi = \pi_1 + i \pi_2$. Since the fields satisfy first order equations,
$\pi$ must contain only annihilation operators and $\pi^\dag$ only creation
operators. So we need two NG fields for describing a physical particle (the NG
boson). With (\ref{K47ab}) and \ref{K47abc}) we have recovered the
well known experimental result (see, e.g. Ref.~\cite{Ashcroft:1976})
that the dispersion relation of ferromagnetic spin waves
has a non-relativistic form. Note that Berry--Anandan
phase was essential in obtaining the right dispersion relation.

The functional integral (\ref{appK.25aa}) with the action
(\ref{K42aa}) represents a particular class of  non-linar $\sigma$
models known as  Landau--Lifshitz $\sigma$ models.
In general, the Landau--Lifshitz $\sigma$ models
are models defined on a  general coset space
$G/H$, with $H$ a maximal stability sub-group of $G$. These are
{\em non-relativistic} models that have $G$-valued Noether charges, local
$H$ invariance and are classically integrable.

Similar analysis can be performed also for
anti-ferromagnets, e.g., along the lines proposed in Ref.~\cite{Fradkin:1988a}.
In this case the classical lowest
energy configuration is described by the N\'{e}el state~\cite{Ashcroft:1976} where the
neighboring lattice spins flip the sign, i.e. ${\bf n}(l) \mapsto
(-1)^l {\bf n}(l)$. The result of absorbing this sign flip is that $H({\bf x}, \dot{\bf x}, t)$
and every other $S_{WZ}[{\bf n}({\bf x})]$ (i.e., Wess--Zumino term of the individual spins) change sign.
With this one can show that the dispersion relation of spin
waves have the linear (relativistic-like) form $\omega({\bf k}) \propto
|{\bf k}|$. This linear gapless dispersion describes the relativistic-like Nambu--Goldstone modes, which are in this case
called {\em anti-ferromagnetic magnons}. It is interesting to point out that in anti-ferromagnets
the corresponding Berry--Anandan phase does not play a dynamical role because in the
N\'{e}el state the Wess--Zumino term reduces to a topological charge~\cite{Fradkin:1988a}.


\section{Final notes\label{final}}

Let us end up with a few notes concerning the presented approach. We have shown that the functional integrals for $G/H\!-\!\sigma$
models which account for quantum dynamics of NG bosons  (i.e., gapless excitations that live in the broken phase of spontaneously broken systems) can be naturally phrased in terms of generalized CS functional integrals. As we have seen, this is because the NG fields take their values in the target space which is the group quotient space $G/H$. Group $G$ in the question is the symmetry of the original (disordered) phase, while $H$ is the residual symmetry after spontaneous symmetry breakdown. State vectors that characterize such NG excitations are then inevitably labeled by points from $G/H$. With a suitable choice of a fiducial state they can be identified with a group-$G$ related CS.

An interesting byproduct of the CS functional integrals is that they naturally generate a Berry--Anandan phase. From
Eq.~(\ref{Eq.26a}) we have seen that the Berry--Anandan phase is determined by the overlaps, i.e., by the inner products, between
CS. Essential in this case is that representations of CS are square-integrable.
Mathematically the Berry--Anandan phase represents anholonomy with respect to the natural (Berry's) connection along a closed loop in the projective Hilbert space~\cite{Anandan}. For CS such a non-trivial anholonomy reflects the ``frustration" of assigning a common phase to all of CS along a closed path in a parameter space~\cite{BJV,Fradkin:1988a}. Closed paths in a parameter space appear typically in the formulation of the partition function. In cases when transition amplitudes are
considered one should work with Pancharatnam's phase instead~\cite{BJ}.  Since the Berry--Anandan phase enters into the action of the CS functional integral it might affect the dynamical properties of the system. In particular, it can (and often it does) change dynamical equations and dispersion relations of the associated NG excitations.


We have illustrated the aforementioned connection between non-linear $\sigma$ models and group-related CS
with a spin-$j$ Heisenberg ferromagent in a broken phase. Apart from the correct dynamical Landau--Lifshitz equations
for quantum ferromagnet we have obtained also correct linear dispersion relation for ferromagnetic magnons.
This was possible only because the Berry--Anandan phase exemplified via Wess--Zumino term furnished the dynamical equations with the first time-derivative term.
It should be further noted that the exact form of the dispersion relation could
not be specified by Goldstone's theorem alone. Dispersion relations are not
determined merely by symmetry considerations, they also crucially depend on the
specifics of the system, namely on the choice of the Hamiltonian $H({\bf x},
\dot{\bf x}, t)$ which specifies the actual interaction between NG fields and on
the spin orientations in respective sublattices which determines type of spin
waves (ferromagnetic or anti-ferromagnetics) and hence type of the NG field. It
is also important to observe that even if we have the same symmetry breaking
pattern $SU(2)\!\rightarrow\!U(1)$, the ferromagnetic and anti-ferromagnetic systems
differ in their qualitative description of the dispersion relation.
For instance, the number of independent magnon  states differs~\cite{Ashcroft:1976};
one for a ferromagnet and two for anti-ferromagnet.
In fact, only the number of real NG fields turns out to be universal and equals to the dimension of the coset space $SU(2)/U(1)$ which is $\mbox{dim}[SU(2)] - \mbox{dim}[U(1)] = 2$ (for ferromagnets these are fields $\pi_1$ and $\pi_2$).

Let us also note that in the large $j$ limit is the $SU(2)$ CS functional integral dominated
by the stationary points of $S_{\rm tot}[{\bf n}]$, i.e. by solutions
of Eq.~(\ref{K46aa}). In fact, with increasing $j$ will the semiclassical representation
of the above $SU(2)$ CS functional integral approximate the exact partition function.
For this reason one might arrange the semiclassical result as power series in $1/j$ in much the same way as the $1/N $ perturbation expansion is done, e.g., in $O(N)$ symmetric models. Such an expansion is known as the Holstein--Primakoff expansion~\cite{Holstein:1940zp}.



\section*{Acknowledgments}

A particular thanks go
to G.~Vitiello, H.~Kleinert and J.~Tolar for enlightening discussions, and to T.W.B.~Kibble, J.~Klauder
and G.~Junker for their constructive feedbacks.
M.B. is supported by
MIUR and INFN.
P.J. is supported by the Ministry of Education of the Czech Republic
under the Grant No. CFRJS 1507001.

\section*{References}

\end{document}